\documentclass[useAMS,usenatbib]{mn2e}
\usepackage{amsmath}
\usepackage{txfonts}
\usepackage{graphics}
\usepackage{times}

\renewcommand{\vec}[1]{\mathbf{#1}}
\newcommand{\vr}{\vec{r}}
\newcommand{\vq}{\vec{q}}

\newcommand{\vk}{\vec{k}}
\newcommand{\vPsi}{\vec{\Psi}}
\newcommand{\vDelta}{\vec{\Delta}}
\newcommand{\hq}{\hat{q}}

\newcommand{\hz}{\hat{z}}

\newcommand{\lam}{\lambda}

\begin{document}


\title[Reconstruction within the Zeldovich approximation]{Reconstruction
within the Zeldovich approximation}
\author[White]{
    Martin White$^{1,2}$ \\
    $^{1}$ Departments of Physics and Astronomy, University of California,
    Berkeley, CA 94720, USA \\
    $^{2}$ Lawrence Berkeley National Laboratory, 1 Cyclotron Road,
    Berkeley, CA 94720, USA
}
\date{\today}
\pagerange{\pageref{firstpage}--\pageref{lastpage}}

\maketitle

\label{firstpage}

\begin{abstract}
The Zeldovich approximation, $1^{\rm st}$ order Lagrangian perturbation
theory, provides a good description of the clustering of matter and galaxies
on large scales.  The acoustic feature in the large-scale correlation
function of galaxies imprinted by sound waves in the early Universe has been
successfully used as a `standard ruler' to constrain the expansion history of
the Universe.  The standard ruler can be improved if a process known as
density field reconstruction is employed.  In this paper we develop the
Zeldovich formalism to compute the correlation function of biased tracers
in both real- and redshift-space using the simplest reconstruction algorithm
with a Gaussian kernel and compare to N-body simulations.
The model qualitatively describes the effects of reconstruction on the
simulations, though its quantitative success depends upon how redshift-space
distortions are handled in the reconstruction algorithm.
\end{abstract}



\section{Introduction}
\label{sec:intro}

The large-scale structure seen in the distribution of galaxies contains
a wealth of information about the nature and constituents of our Universe.
Of particular interest here is the use of low-order statistics of this field
to constrain the distance scale and growth rate of fluctuations, which in
turn impact upon our understanding of dark energy and tests of General
Relativity at cosmological scales \citep[e.g.][]{PDG}.
One of the premier methods for measuring the distance scale\footnote{And for
breaking degeneracies when constraining parameters from the cosmic microwave
background anisotropies, e.g.~\citet{PlanckA15}.}
uses the baryon acoustic oscillation (BAO) `feature' in the 2-point function
of galaxies as a calibrated, standard ruler
\citep[see][for a review]{PDG}.
Additional information on the rate of growth of perturbations, which allows a
key test of General Relativity and constraints on modified gravity
\citep[e.g.][and references therein]{Joy14},
is encoded in the anisotropy of the 2-point function imprinted by peculiar
velocities, i.e.~redshift space distortions \citep[see][for a review]{Ham98}.
Fits to the distance scale using the BAO feature become significantly more
accurate if density field `reconstruction' is applied \citep{Eis07}, but
this procedure alters the signal that is used to infer the growth rate from
redshift-space distortions.
Ideally we would have a model which can simultaneously describe the features
which are used to constrain distance scale and the growth of structure, since
there is a non-trivial degeneracy between mis-estimates of distance and
growth \citep[e.g.~Fig.~9 of][]{Rei12}.
A formalism which can be used to simultaneously describe both of these
pieces of a redshift survey is currently not known.

It is straightforward to form a data vector which consists of the correlation
function pre-reconstruction on small scales and post-reconstruction on large
scales.
Our goal is to find a single theoretical framework which could simultaneously
fit both parts of this data vector\footnote{Obviously, such a model would also
form a good template for fitting the BAO peak position on its own.}.
Models based upon Lagrangian perturbation theory have been shown to do a
good job of fitting the anisotropic signal in the (pre-reconstruction)
correlation function (see e.g.~\citealt{Whi15} for a recent investigation
and references to the earlier literature).
In this paper we investigate how accurately $1^{\rm st}$ order Lagrangian
perturbation theory (``the Zeldovich approximation'') can be used to model
the reconstructed BAO feature in the redshift-space correlation function of
biased tracers.

The last few years have seen a resurgence of interest in the Zeldovich
approximation.  It has been applied to understanding the effects of non-linear
structure formation on the baryon acoustic oscillation feature in the
correlation function \citep{PadWhi09,McCSza12,TasZal12a} and to
understanding how ``reconstruction'' \citep{Eis07} removes those
non-linearities \citep{PadWhiCoh09,NohWhiPad09,TasZal12b}.
It has been used as the basis for an effective field theory of large-scale
structure \citep{PorSenZal14} and a new version of the halo model
\citep{SelVla15}.
It has been compared to ``standard'' perturbation theory \citep{Tas14a},
extended to higher orders in Lagrangian perturbation theory
\citep{Mat08a,Mat08b,OkaTarMat11,CLPT,VlaSelBal15}
and to higher order statistics \citep{Tas14b} including a model for the
power spectrum covariance matrix \citep{MohSel14}.
Despite the more than 40 years since it was introduced, the Zeldovich
approximation still provides one of our most accurate models for the
distribution of cosmological objects.

The outline of the paper is as follows.
Section \ref{sec:review} contains a review of the salient aspects of
Lagrangian perturbation theory and reconstruction, to fix our notation,
and introduces our N-body simulations.
Section \ref{sec:zeldovichRecon} introduces the Zeldovich model for
reconstruction and compares its predictions to the simulations.
We finish in Section \ref{sec:discussion} with an assessment of the
Zeldovich approximation and future directions for research.

\section{Background and review}
\label{sec:review}

\subsection{Lagrangian perturbation theory}

We wish to develop an analytic description of the reconstructed correlation
function of biased tracers in redshift space and to this end we use Lagrangian
perturbation theory\footnote{See \citet{Ber02} for a comprehensive (though
somewhat dated) review of Eulerian perturbation theory.}
\citep{Buc89,Mou91,Hiv95,TayHam96}.
In this section we remind the reader of some essential terminology, and
establish our notational conventions.
Our notation and formalism follows closely that in
\citet{Mat08a,Mat08b,CLPT,WanReiWhi13,Whi14}
to which we refer the reader for further details and original references.

In the Lagrangian approach to cosmological fluid dynamics, one traces the
trajectory of an individual fluid element through space and time.
Every element of the fluid is uniquely labeled by its Lagrangian coordinate
$\vq$ and the displacement field $\vPsi(\vq,t)$ fully specifies the motion
of the cosmological fluid.  Lagrangian Perturbation Theory (LPT) develops a
perturbative solution for $\vPsi$ but we shall deal here with the first
order solution which is known as the Zeldovich approximation \citep{Zel70}.
Denote this first order solution as $\vPsi$ we have:
\begin{equation}
    \vPsi(\vq) = \int \frac{d^3k}{(2\pi)^3}
    \ e^{i\vk\cdot\vq} \frac{i\vk}{k^2} \delta_L(\vk) ,
\end{equation}
We shall assume that halos, and the galaxies that inhabit them, have a
local Lagrangian bias $\rho_X(\vq) = \bar{\rho}_X F[\delta_R(\vq)]$.
\citet{Mat11} provides an extensive discussion of local and non-local
Lagrangian bias schemes.

This formalism makes it particularly easy to include redshift space
distortions.  We follow the earlier papers and adopt the ``plane-parallel''
or ``distant-observer'' approximation, in which the line-of-sight direction
to each object is taken to be the fixed direction $\hat{z}$.
Within this approximation, including redshift-space distortions is
achieved via
\begin{equation}
 \Psi_i \to \Psi_i^s = R_{ij}\Psi_j = (\delta_{ij} + f \hz_i \hz_j) \Psi_j
\label{eqn:redshiftspace}
\end{equation}
which simply multiplies the $z$-component of the vector by $1+f$.

The correlation function within the Zeldovich approximation then follows
by elementary manipulations.
Defining $\Delta \equiv \vPsi_2 - \vPsi_1$ and writing $F_i=F(\lambda_i)$
for the Fourier transform of $F[\delta_R(\mathbf{q})]$ the real-space
correlation function is
\begin{eqnarray}
    1 + \xi_X(\vr) &=& \int d^3q
    \int \frac{d^3k}{(2\pi)^3} e^{i\vk\cdot(\vq-\vr)}
    \int \frac{d\lam_1}{2\pi} \frac{d\lam_2}{2\pi}
    \ F_1 F_2 \nonumber \\
    &\times& \left\langle
    e^{i(\lam_1 \delta_1 + \lam_2 \delta_2 + \vk\cdot\vDelta)} \right\rangle .
\end{eqnarray}
For convenience we define
$\xi_L(\vq) = \langle \delta_1 \delta_2 \rangle$,
$U_i(\vq) = \langle \delta_1 \Delta_i \rangle 
          = \langle \delta_2 \Delta_i \rangle$,
and $A_{ij}(\vq)= \langle \Delta_i \Delta_j \rangle$.
The vector $U_i(\vq)=U(q)\,\hq_i$ is the cross-correlation between the linear
density field and the Lagrangian displacement field.
The matrix $A_{ij}$ may be decomposed as
\begin{eqnarray}
  A_{ij}(\vq) &=& 2\left[\sigma_\eta^2 - \eta_\perp(q)\right] \delta_{ij}
               + 2\left[\eta_\perp(q) - \eta_\parallel(q)\right] \hq_i \hq_j ,\\
  &=& \sigma_\perp^2\delta_{ij}
   + \left[\sigma_\parallel^2-\sigma_\perp^2\right]\hq_i\hq_j
\label{eqn:Aijdef}
\end{eqnarray}
where $\sigma_\eta^2 \equiv \frac{1}{3} \langle |\vPsi|^2 \rangle$ is the 1-D
dispersion of the displacement field, and $\eta_\parallel$ and $\eta_\perp$ are
the transverse and longitudinal components of the Lagrangian 2-point function,
$\eta_{ij}(\vq) = \left\langle \Psi_i(\vq_1) \Psi_j(\vq_2) \right\rangle$.
In the Zeldovich approximation these quantities are given by simple integrals
over the linear power spectrum:
\begin{gather}
  \sigma_\eta^2 = \frac{1}{6\pi^2} \int_0^\infty dk~ P_L(k) , \\
  \eta_\perp(q) = \frac{1}{2\pi^2} \int_0^\infty dk~ P_L(k)~
  \frac{j_1(kq)}{kq} , \\
  \eta_\parallel(q) = \frac{1}{2\pi^2} \int_0^\infty dk~ P_L(k)~
  \left[j_0(kq) - 2 \frac{j_1(kq)}{kq}\right] , \\
  U(q) = -\frac{1}{2\pi^2} \int_0^\infty dk~ k P_L(k)~ j_1(kq) .
\label{eqn:qf}
\end{gather}

Taylor series expanding the bias terms and doing the $\lam_1$ and
$\lam_2$ integrations and the Fourier transform we can write
\begin{eqnarray}
  1 + \xi_X(\vr) &=& \int \frac{d^3q}{(2\pi)^{3/2} |A|^{1/2}}
  \ e^{-\frac{1}{2} (\vr-\vq)^T \mathbf{A}^{-1} (\vr-\vq)}
  \left[ 1 + b_1^2 \xi_L \right. \nonumber \\
  &-& 2 b_1 U_i g_i + \frac{1}{2} b_2^2 \xi_L^2
   - (b_2 + b_1^2) U_i U_j G_{ij} \nonumber \\
  &-& \left. 2b_1 b_2 \xi_L U_i g_i + \cdots \right] ,
\label{eqn:xi_ZA}
\end{eqnarray}
where we have written $b_n=\left\langle F^{(n)}\right\rangle$,
$g_i\equiv (A^{-1})_{ij}(q-r)_j$ and
$G_{ij}\equiv (A^{-1})_{ij} - g_i g_j$
in order to make the expressions more readable.
The generalization to redshift space follows straightforwardly from
Eq.~(\ref{eqn:redshiftspace}): we simply multiply $U_z$ by $1+f$ and divide
the $z$-components of $\mathbf{A}^{-1}$ by the same factor.

Not all of the terms in Eq.~(\ref{eqn:xi_ZA}) are important at the scales
relevant for BAO.
For typical values of halo bias ($b_1\sim 1$ and $b_2\sim 0.1$), the dominant
contributions to the real space correlation function or the monopole of the
redshift space correlation function at $r\simeq 100\,h^{-1}$Mpc are from
the ``1'', $b_1^2\xi_L$ and $-2b_1U_ig_i$ terms.
The other terms make up less than one per cent of the total.
For the quadrupole of the redshift space correlation function only the
the ``1'' and $-2b_1U_ig_i$ terms contribute significantly
\citep[see also][Fig.~4]{Whi14}.

\subsection{Reconstruction}
\label{sec:recon}

We start by reviewing the reconstruction algorithm of \citet{Eis07}
and its interpretation within Lagrangian perturbation theory
\citep{PadWhiCoh09,NohWhiPad09}.
Various tests of reconstruction have been performed in
\citet{Seo10,Pad12,Xu13,Bur14,Toj14}
which also contain useful details on the specific implementations.

The algorithm devised by \citet{Eis07} is straightforward to apply
and consists of the following steps:
\begin{itemize}
\item Smooth the halo or galaxy density field with a kernel
$\mathcal{S}$ (see below) to filter out small scale (high $k$) modes,
which are difficult to model.  Divide the amplitude of the overdensity
by an estimate of the large-scale bias, $b$, to obtain a proxy for the
overdensity field: $\delta(\mathbf{x})$.
\item Compute the shift, $\mathbf{s}$, from the smoothed density field in
redshift space using the Zeldovich approximation
(this field obeys $\mathbf{\nabla}\cdot\mathbf{R}\mathbf{s}=-\delta$ with
 the $f$ replaced by $f/b$ in $\mathbf{R}$).
The line-of-sight component of $\mathbf{s}$ is multiplied by $1+f$ to
approximately account for redshift-space distortions.
\item Move the galaxies by $\mathbf{s}$ and compute the ``displaced''
density field, $\delta_d$.
\item Shift an initially spatially uniform distribution of particles by
$\mathbf{s}$ to form the ``shifted'' density field, $\delta_s$.  It is
ambiguous whether this shift includes the factor of $1+f$ in the line-of-sight
direction or not.  Including the $1+f$ includes `linear' redshift-space
distortions in the reconstructed field while excluding it removed them.
\citet{Pad12,Xu13} and later works do not include this factor, but earlier
papers did not distinguish between the uniform sample and the galaxies.
We shall consider both approaches.
\item The reconstructed density field is defined as
$\delta_r\equiv \delta_d-\delta_s$ with power spectrum
$P_r(k)\propto \langle \left| \delta_r^2\right|\rangle$.
\end{itemize}
Following \citet{Eis07} we use a Gaussian smoothing of scale $R$,
specifically $\mathcal{S}(k) =  e^{-(kR)^2/2}$.
Throughout we shall assume that the fiducial cosmology, bias and $f$ are
properly known during reconstruction.
\citet{Pad12,Xu13,Bur14,Var14} show that the reconstructed 2-point function is
quite insensitive to the specific choices made, so this is a reasonable first
approximation.
We shall return to this issue in Section \ref{sec:discussion}.

\subsection{N-body simulations}

We use a suite of 20 N-body simulations to test how well the Zeldovich
model works.  The simulations assume a $\Lambda$CDM cosmology with
$\Omega_m = 0.274$, $\Omega_\Lambda = 0.726$, $h = 0.7$, $n = 0.95$, and
$\sigma_8 = 0.8$ and were run with the TreePM code described in \citet{TreePM}.
Each simulation employed $1500^3$ equal mass
($m_p\simeq 7.6\times10^{10}\,h^{-1}M_\odot$)
particles in a periodic cube of side length $1.5\,h^{-1}$Gpc
as described in \citet{ReiWhi11} and \citet{Whi11}.
Halos are found using the friends-of-friends method, with a linking length
of $0.168$ times the mean inter-particle spacing.
These are the same simulations and catalogs that were used in
\citet{WanReiWhi13,Whi14,Whi15}
and further details can be found in those papers.
Throughout we shall use halos with friends-of-friends mass in the range
$12.785<\log_{10}M_h/(h^{-1}M_\odot)<13.085$, with $b\simeq 1.7$, which
is one of the samples used in \citet{WanReiWhi13,Whi14}.  It has a
relatively high bias, while at the same time a large enough spatial
density to reduce shot noise to tolerable levels.

\section{Zeldovich reconstructed}
\label{sec:zeldovichRecon}

With this background in hand it is now straightforward to develop a
model for the reconstructed correlation function within the Zeldovich
approximation.

\subsection{The shift}

\begin{figure}
\begin{center}
\resizebox{\columnwidth}{!}{\includegraphics{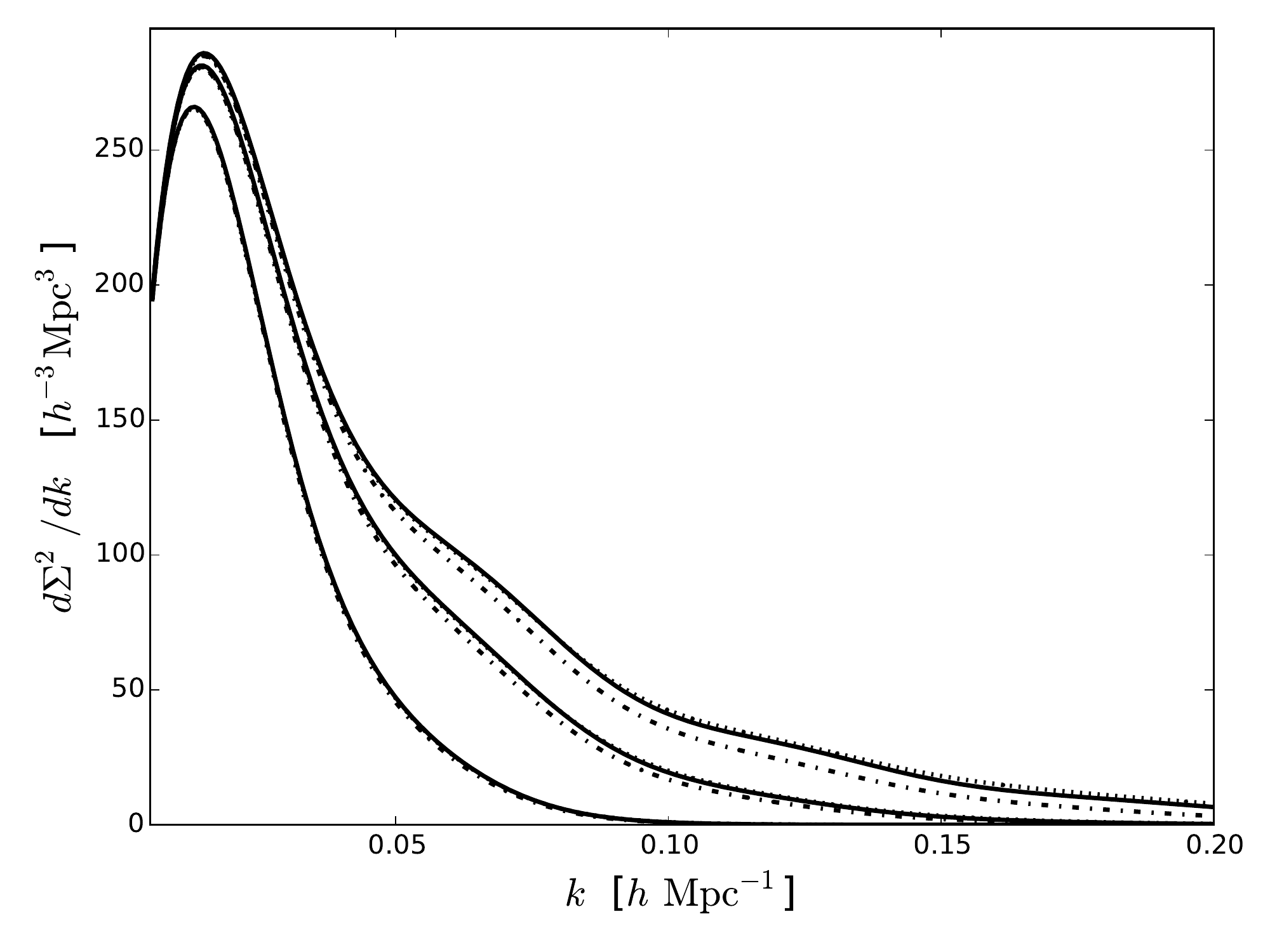}}
\end{center}
\caption{The contribution to the variance of the 1D Zeldovich displacement,
per unit $k$, at $z\simeq 0.5$ for three different (Gaussian) smoothing scales:
$R=5$, 10 and $20\,h^{-1}$Mpc (upper to lower sets of lines).
For each set of lines the solid line is the linear theory prediction, the
dashed line assumes standard, $2^{\rm nd}$ order, Eulerian perturbation theory
and the dot-dashed line is the Zeldovich approximation expanded to $2^{\rm nd}$
order.  Except for the $5\,h^{-1}$Mpc case all three approximations are in
excellent agreement (see also Fig.~1 of \citealt{PadWhiCoh09}).}
\label{fig:sigma}
\end{figure}

We will assume that the ``shift'' field, which is formally computed on the
non-linear density field at the Eulerian position, $\mathbf{x}$, can be
well approximated by the negative Zeldovich displacement computed from the
linear theory field at the Lagrangian position, $\mathbf{q}$.
This is a reasonable first approximation since such shifts are dominated
by very long wavelength modes \citep{ESW07}.
The difference between $\delta_L(\mathbf{q})$ and $\delta_L(\mathbf{x})$
is higher-order in $\mathbf{\Psi}$ and so should be comparable to the
effect of non-linearities in the density\footnote{While the `shifts' from
Lagrangian to Eulerian coordinates are large in CDM, they are quite coherent
so this approximation is not as drastic as it at first seems.}.
Within the same approximation, solving
$\mathbf{\nabla}\cdot\mathbf{R}\mathbf{s}=-\delta$
on the redshift-space field is the same as generating
$\mathbf{s}(\mathbf{k})=-i(\mathbf{k}/k^2)\delta(\mathbf{k}) \mathcal{S}(k)$
using the real-space field.

To estimate the relative size of the correction to the shift terms coming from
non-linearities in the density, we look at the contributions to the rms
Zeldovich displacement for different (Gaussian) smoothing scales, $R$.
In real space the 1D displacement is
$\left[\int dk\, P(k)/(6\pi^2)\right]^{1/2}$.
Fig.~\ref{fig:sigma} shows the fractional contribution to the squared
displacement from beyond-linear terms in $P(k)$, computed from (standard)
Eulerian perturbation theory or the Zeldovich approximation
[see Appendix \ref{app:pk} for more details].
For smoothings of $10\,h^{-1}$Mpc or above the approximation appears to be
very good.  We shall use $R=15\,h^{-1}$Mpc as our default
\citep[as used in e.g.][]{Pad12,And14,Toj14}, unless otherwise specified.

Under this approximation we compute the statistics of the displaced field
by replacing $\vPsi$ with $\vPsi+\vec{s}$ and of the shifted field by
replacing $\vPsi$ with $\vec{s}$ in the formulae of \S\ref{sec:review}.

\subsection{Real space}

\begin{figure}
\begin{center}
\resizebox{\columnwidth}{!}{\includegraphics{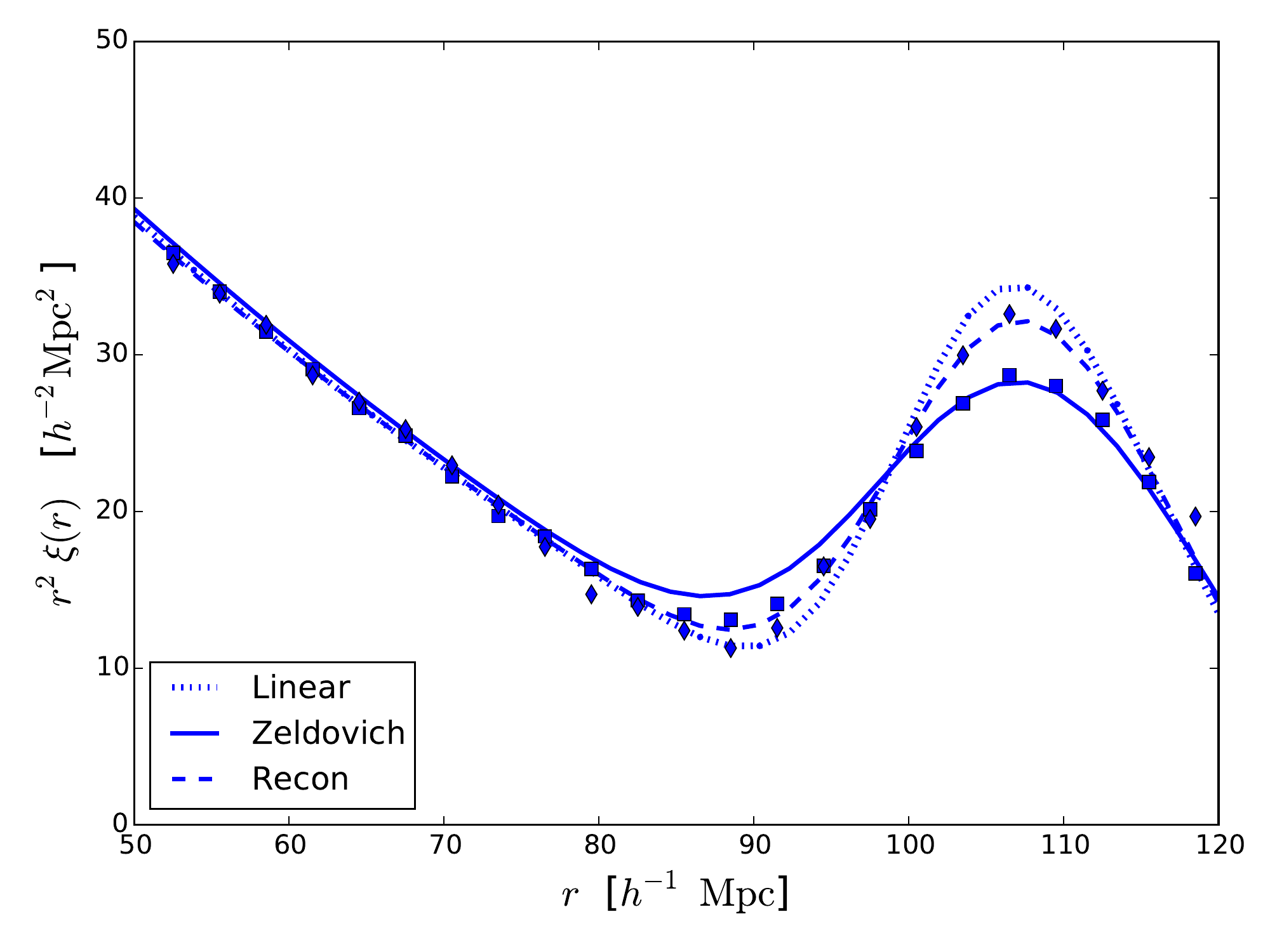}}
\end{center}
\caption{A comparison of the Zeldovich and N-body real-space, halo
correlation functions pre- and post-reconstruction.  The dotted
line shows the linear theory, while the solid (dashed) line shows the
Zeldovich prediction pre(post)-reconstruction.  The squares and diamonds
show the unreconstructed and reconstructed results from the N-body
simulations described in the text.
We have used a smoothing scale of $R=15\,h^{-1}$Mpc when performing
reconstruction.}
\label{fig:zelrecon_real}
\end{figure}

Let us first consider the statistics of the reconstructed field in real
space.  The reconstructed field is the sum of the displaced and the negative
of the shifted fields of Sec.~\ref{sec:recon} and thus the correlation function
has 3 terms: the auto-correlation of the displaced field, the auto-correlation
of the negative-shifted field and the cross-correlation of the two fields:
$\xi^{(\rm recon)} = \xi^{(dd)} + \xi^{(ss)} + 2\xi^{(ds)}$.
Each term will have the same functional form as Eq.~(\ref{eqn:xi_ZA}).
Let us take each in turn.
The auto-correlation function of the displaced field, $\xi^{(dd)}$, is given
by Eq.~(\ref{eqn:xi_ZA}) with $P_L\to P_L(1-\mathcal{S})^2$ when evaluating
$\eta_\perp$ and $\eta_\parallel$ and one power of $1-\mathcal{S}$ when
computing $U$ (it is unchanged when computing $\xi_L$).  Thus for example
the $U_i$ entering the analog of Eq.~(\ref{eqn:xi_ZA}) for $\xi^{(dd)}$ is
given by
\begin{equation}
  U^{(dd)}(q) = -\frac{1}{2\pi^2}\int_0^\infty dk\,k
  \ P_L(k)\left(1-\mathcal{S}\right)\ j_1(kq) .
\end{equation}
and similarly for the other terms.
The auto-correlation function of the shifted field is similarly given by
Eq.~(\ref{eqn:xi_ZA}) with $b_1=b_2=0$ (i.e.~the terms in square brackets
in Eq.~(\ref{eqn:xi_ZA}) become $1$) and $P_L\to P_L\mathcal{S}^2$ when
evaluating $\eta_\perp$ and $\eta_\parallel$ which define $A_{ij}$.
The cross term between the displaced and shifted fields has
$P_L\to P_L\mathcal{S}(1-\mathcal{S})$ when evaluating $\eta_\perp$ and
$\eta_\parallel$ and $P_L\to P_L\mathcal{S}$ when evaluating $U$ and the
substitutions $b_1\to\frac{1}{2}b_1$, $b_2\to\frac{1}{2}b_2$, $b_1^2\to 0$,
$b_2^2\to 0$ and $b_1b_2\to 0$ in Eq.~(\ref{eqn:xi_ZA}), i.e.
\begin{eqnarray}
  1 + \xi_X^{(ds)}(\vr) &=& \int \frac{d^3q}{(2\pi)^{3/2} |A^{(ds)}|^{1/2}}
  \ e^{-\frac{1}{2} (\vr-\vq)^T \mathbf{A}_{(ds)}^{-1} (\vr-\vq)}
  \left[ 1 \right. \nonumber \\
  && \left. - b_1 U_i^{(ds)} g_i^{(ds)} -
     \frac{1}{2}b_2 U_i^{(ds)} U_j^{(ds)} G_{ij}^{(ds)} + \cdots\right]
\end{eqnarray}

A comparison of the correlation function predicted by the Zeldovich
approximation with that measured in N-body simulations is shown in
Fig.~\ref{fig:zelrecon_real}.  The theory predicts that the acoustic
peak (at $r\simeq 110\,h^{-1}$Mpc) is broadened by the effects of non-linear
structure formation and that reconstruction acts to sharpen the peak.
The agreement with the simulations both pre- and post-reconstruction is quite
good, as expected from the earlier work of
\citet[][although in that work $2^{\rm nd}$ order LPT was used]{NohWhiPad09}.
While we do not have the necessary volume of simulations to reliably measure
the peak location at sub-percent precision, we argue in the Appendix that
the model should accurately reflect the manner in which reconstruction reduces
the small shift in the peak location engendered by mode-coupling
(see similar discussion in \citealt{PadWhi09}).
We have checked that the agreement between the model and the simulations is
qualitatively similar for variations in the smoothing scale between
$10$ to $20\,h^{-1}$Mpc.

\subsection{Redshift space}

\begin{figure}
\begin{center}
\resizebox{\columnwidth}{!}{\includegraphics{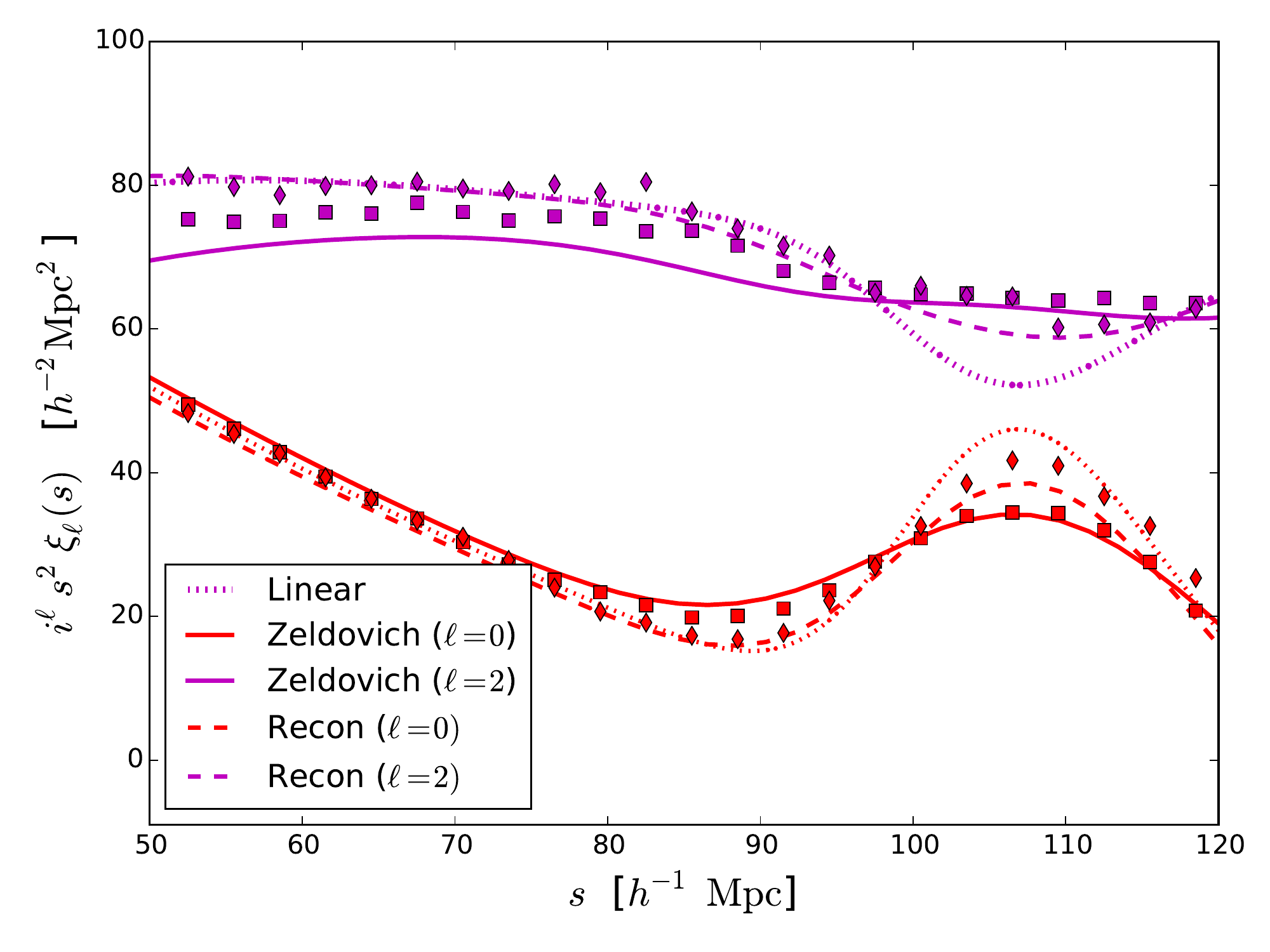}}
\resizebox{\columnwidth}{!}{\includegraphics{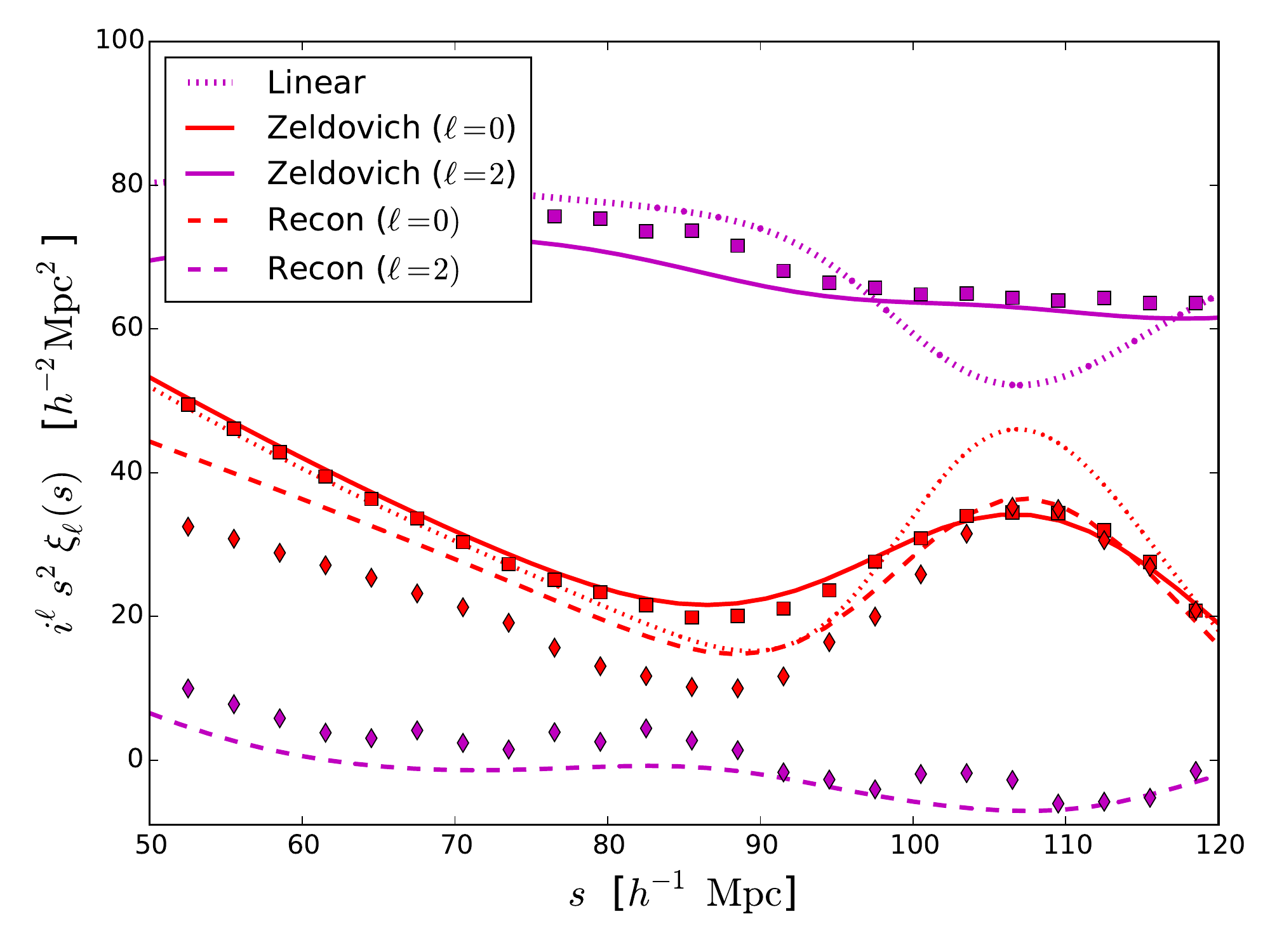}}
\end{center}
\caption{A comparison of the Zeldovich and N-body redshift-space, halo
correlation functions pre- and post-reconstruction.  The dotted
line shows the linear theory, while the solid (dashed) line shows the
Zeldovich prediction pre(post)-reconstruction.  The squares and diamonds
show the unreconstructed and reconstructed results from the N-body
simulations described in the text.
The upper set of lines are for the quadrupole while the lower set of lines
is for the monopole except in the lower panel where the lowest dashed line
is for the reconstructed quadrupole.
Two versions of reconstruction are shown: (upper) with both the halos and
the initially uniformly distributed particles shifted by the same field
(lower) with the halos shifted $1+f$ times further in the line-of-sight
direction than the uniform particles.}
\label{fig:zelrecon_red}
\end{figure}

Now we turn to redshift space.
If we use a single field, $\mathbf{s}$, to shift both the halos and the
random particles (i.e.~with the factor of $1+f$ in the line-of-sight
direction for both) when generating $\vec{s}$ the modifications to the
preceeding section are small: we simply multiply $U_z$ by $1+f$ and divide
the $z$-components of $\mathbf{A}^{-1}$ by the same factor.

The upper panel of Fig.~\ref{fig:zelrecon_red} shows the monopole and
quadrupole of the correlation function in this case.
The Zeldovich approximation does a credible job of fitting the monopole
of the redshift-space, halo correlation function pre-reconstruction.
The agreement for the quadrupole moment is better than linear theory in the
acoustic peak region, but not as good as for the monopole
(as expected from earlier work, e.g.~\citealt{Whi14}, Fig.~2).
To avoid cluttering the figure we have not plotted the errors on the N-body
points.  For the monopole they are generally small, but for the quadrupole
(pre- and post-reconstruction) they are significant.
In the acoustic peak region the typical error on $s^2\xi_2$ is
$3-5\,h^{-2}{\rm Mpc}^2$ and the errors are highly correlated.
Post-reconstruction the results for both multipoles of the correlation
function are qualitatively similar: the reconstructed multipoles are closer
to the linear theory than the evolved ones and the agreement with the
N-body simulations in the region of the acoustic peak
($s\simeq 110\,h^{-1}$Mpc) is quite good.
Unfortunately the errors on the quadrupole from the N-body simulations are
too large to see whether the predicted shift from the pre- to
post-reconstruction shape near the acoustic peak is borne out in simulations.
If pushed to smaller scales the model starts to depart significantly from the
simulation results, no doubt because the Zeldovich approximation does not
accurately capture the anisotropies in the displacement/velocity field on
smaller scales \citep[see][for further discussion]{Whi14}.
There is weak evidence that the Zeldovich approximation agrees better with
the N-body simulations for the quadrupole moment after reconstruction than
it does before.
Increasing the smoothing scale (to $30\,h^{-1}$Mpc) leads to similar
agreement between the simulation and model, but reduces the sharpening of
the peak by reconstruction.  Reducing the smoothing scale to $10\,h^{-1}$Mpc
gives results very similar to those shown in Fig.~\ref{fig:zelrecon_red}.

An alternative formulation does not include the factor of $1+f$ in the
line-of-sight shift for the initially uniformly distributed particles.
This acts to reduce the effects of redshift-space distortions in the
reconstructed density field.
In this case the factors of $1+f$ are omitted entirely when computing the
shift-shift auto-correlation function, and only one power of $1+f$ is
included in $\mathbf{A}^{-1}$ and no factors of $1+f$ in $\mathbf{U}$ in
the cross-correlation of the displaced and shifted particles but the rest
of the terms remain unchanged.
This is shown in the lower panel of Fig.~\ref{fig:zelrecon_red} and the
level of agreement between the theory and the simulations is similar
to that in the upper panel.
Note in the lower panel the quadrupole is significantly reduced in both the
model and the simulations, indicating that we have removed most of the effects
of linear redshift-space distortions, but it is not reduced entirely to zero
(earlier investigations of reconstruction in simulations either did not include
redshift-space distortions or presented only the monopole statistics).
Again the numerical errors from the N-body simulation are not negligible, but
the overall trends are clear.  The agreement between the simulations and the
model in the monopole is no longer as good on scales smaller than the acoustic
peak as it was in the upper panel.

Comparing the upper and lower panels of Fig.~\ref{fig:zelrecon_red} suggests
that the errors in how the Zeldovich approximation models reconstruction
partially cancel if both the galaxies and initially uniformly distributed
sample of particles are shifted by the same field.  In this case the agreement
between the model and simulations in both the monopole and quadrupole moments
of the correlation function above $90\,h^{-1}$Mpc is quite encouraging.
If only the galaxies are shifted by an additional factor of $1+f$ in the
line-of-sight direction the reduction in the quadrupole moment is qualitatively
reproduced by the model but the well-known inaccuracies in the halo velocity
field cause a significant over-estimate of the monopole even at
$90\,h^{-1}$Mpc.  If the Zeldovich approximation is to be used as a template
for fitting the reconstructed BAO feature, it would be better to implement
reconstruction on the data using the `both shift' formalism.
If the behavior of the model is improved because the `both shift' formulation
reduces sensitivity to small scales (where the model does less well) then this
formulation may be less sensitive to small scales in the data as well and
potentially more robust.  Such an investigation is outside the scope of this
work.

\section{Discussion}
\label{sec:discussion}

The goal of this paper was to investigate a model for the reconstructed,
redshift-space correlation function of biased tracers within the framework
of Lagrangian perturbation theory.  In principle such a model can be
combined with other models within the same framework to fit a combination
of data such as reconstructed BAO and redshift-space distortions, for
example by fitting a data vector which consists of pre-reconstruction
multipoles below $s\simeq 90\,h^{-1}$Mpc and reconstructed multipoles above
$s\simeq 90\,h^{-1}$Mpc.

Previous work \citep{PadWhiCoh09,NohWhiPad09} developed the iPT formalism
of \citet{Mat08a,Mat08b} to reconstruction in real space and made comparison
to N-body simulations.
In this work we have specialized to lowest order in LPT, i.e.~the
Zeldovich approximation, but avoided some of the perturbative expansions
inherent in iPT, extended the model to include redshift-space distortions
and compared to a larger set of N-body simulations.

The Zeldovich model performs very well, in comparison to N-body simulations,
for the real-space correlation function of halos both pre- and
post-reconstruction.  In redshift space the monopole moment of the correlation
function is well reproduced, and the quadrupole moment is consistent near
the acoustic peak.
Post-reconstruction the model correctly reproduces the sharpening of the
acoustic peak and the modification of the quadrupole, but the quantitative
agreement is not as good as in real space.  The range of scales over which
the model and the simulations agree depends upon how the reconstruction
algorithm is implemented, with best agreement if both the `displaced' and
`shifted' fields are shifted by the same amount.

We have concentrated on developing and validating the Zeldovich approximation
for reconstruction, assuming that implementation details, survey non-idealities
and misestimates of the various parameters in reconstruction introduce effects
that are subdominant to the statistical errors.
This is likely true for the current generation of surveys
\citep[e.g.][]{Pad12,And14} but may need to be revised for future surveys.
One possibility is to rerun reconstruction, and recompute the 2-point
statistics, for each cosmology whose likelihood is being evaluated
(in which case the fiducial cosmology, bias and growth factor will be
self-consistently included).
This is extremely expensive, computationally.
For small variations in parameters it may be possible to develop a linear
response model for the 2-point function, or an emulator.
Alternatively, an obvious direction for development is to model misestimates
of $b$, $f$ and the fiducial cosmology within the Zeldovich approximation.
This adds significant complexity to the calculation and obscures the main
points of this paper, but may be a more computationally efficient method of
proceeding when fitting data.  As a side benefit it could allow an analytic
understanding of the manner in which such assumptions impact the inferences.
We defer such development to future work.

\vspace{0.2in}
I would like to thank Shirley Ho for helpful comments on an earlier draft.
This work made extensive use of the NASA Astrophysics Data System and
of the {\tt astro-ph} preprint archive at {\tt arXiv.org}.
The analysis made use of the computing resources of the National Energy
Research Scientific Computing Center.


\newcommand{\aj}{AJ}
\newcommand{\apj}{ApJ}
\newcommand{\apjs}{ApJ Suppl.}
\newcommand{\mnras}{MNRAS}
\newcommand{\araa}{ARA{\&}A}
\newcommand{\aap}{A{\&}A}
\newcommand{\pre}{PRE}
\newcommand{\prd}{Phys. Rev. D}
\newcommand{\apjl}{ApJL}
\newcommand{\physrep}{Physics Reports}
\newcommand{\nat}{Nature}

\setlength{\bibhang}{2.0em}
\setlength\labelwidth{0.0em}
\bibliography{biblio}

\begin{appendix}

\section{Zeldovich vs.~Eulerian PT}
\label{app:pk}

Here we briefly discuss the second order contributions to $P(k)$ in
(standard) Eulerian perturbation theory and in the Zeldovich approximation.
In the latter case it is possible to write down an expression for $P(k)$
to infinite order, but here we shall focus on the $2^{\rm nd}$ order
contributions.

In both cases the second order contribution is the sum of
\begin{equation}
  P^{(2,2)}(k) = 2 \int\frac{d^3p}{(2\pi)^3}
  F_2(\vec{p},\vec{k}-\vec{p}) F_2(-\vec{p},\vec{p}-\vec{k})
  \ P_L(p)P_L(|\vec{k}-\vec{p}|)
\end{equation}
and
\begin{equation}
  P^{(1,3)}(k) = 6P_L(k) \int\frac{d^3p}{(2\pi)^3}
  F_3(\vec{k},\vec{p},-\vec{p})\ P_L(p)
\end{equation}
where $F_n$ are the well-known perturbation theory kernels
\citep[e.g.][]{GGRW,Ber02}.

For the Zeldovich approximation we have \citep{GriWis87}
\begin{equation}
  F_n(\vec{p}_1,\cdots,\vec{p}_n) = \frac{1}{n!} \prod_{i=1}^n
  \frac{\vec{k}\cdot\vec{p}_i}{p_i^2}
\end{equation}
where $\vec{k}=\sum_{i=1}^n \vec{p}_i$.
Thus
\begin{eqnarray}
  F_2(\vec{p},\vec{k}-\vec{p}) F_2(-\vec{p},\vec{p}-\vec{k}) &=&
  \frac{1}{4}\left[
  \frac{\vec{k}\cdot\vec{p}(k^2-\vec{k}\cdot\vec{p})}{p^2|\vec{k}-\vec{p}|^2}
  \right]^2 \\
  &=& \frac{\mu^2(\mu-r)^2}{4(1-2\mu r + r^2)^2}
\label{eqn:F22Z}
\end{eqnarray}
where we have written $\hat{p}\cdot\hat{k}=\mu$ and $p=kr$.
For standard perturbation
theory \citep{Ber02}
\begin{equation}
  F_2(\vec{p}_1,\vec{p}_2) = \frac{5}{7} + \frac{1}{2}
  \frac{\vec{p}_1\cdot\vec{p}_2}{p_1p_2}
  \left(\frac{p_1}{p_2}+\frac{p_2}{p_1}\right) + \frac{2}{7}
  \frac{(\vec{p}_1\cdot\vec{p}_2)^2}{p_1^2p_2^2}
\label{eqn:F2spt}
\end{equation}
thus
\begin{equation}
  F_2(\vec{p},\vec{k}-\vec{p}) F_2(-\vec{p},\vec{p}-\vec{k}) =
  \frac{1}{196} \frac{(7\mu+3r-10\mu^2r)^2}{r^2(1-2\mu r+r^2)^2}
  \quad .
\label{eqn:F22E}
\end{equation}
In both cases the integral over the azimuthal angle is trivial, and
we are left with the $\mu$ and $r$ integrals:
\begin{equation}
  P^{(2,2)}(k) = \frac{k^3}{2\pi^2}\int_0^\infty dr\ P(kr)
  \int_{-1}^{+1}d\mu\ P\left(k\sqrt{1+r^2-2r\mu}\right) F_2^2(r,\mu)
\end{equation}
where we have written $F_2^2(r,\mu)$ as a short-hand for the expressions
in Eqs.~(\ref{eqn:F22Z},\ref{eqn:F22E}).

For $P^{(1,3)}$ we need to evaluate $F_3(\vec{k},\vec{p},-\vec{p})$.
In the Zeldovich approximation we have
\begin{equation}
  F_3(\vec{k},\vec{p},-\vec{p}) = -\frac{1}{3!}\frac{\mu^2}{r^2}
\end{equation}
while for standard perturbation theory the expression involving the
symmetrized form of $F_3$ is quite lengthy and won't be reproduced here.
Performing the azimuthal integral we then obtain the well known result
for $P^{(1,3)}$ in the Zeldovich approximation:
\begin{equation}
  P^{(1,3)}(k) = -k^2 P_L(k) \int_0^\infty\frac{dp}{6\pi^2}\, P_L(p)
\end{equation}
while for standard perturbation theory
\begin{eqnarray}
  P^{(1,3)}(k) &=& \frac{k^3\,P_L(k)}{1008\pi^2} \int_0^\infty dr\ P_L(kr)
  \left[\frac{12}{r^2} - 158 + 100r^2 - 42r^4 \right. \nonumber \\
  &+& \left. \frac{3(r^2-1)^3(7r^2+2)}{r^2}
  \ln\left|\frac{1+r}{1-r}\right|\right]
\end{eqnarray}

It is well established that Lagrangian perturbation theory, and the
Zeldovich approximation, accurately describe the broadening of the acoustic
peak.
At this point it is also straightforward to understand the origin of
``shifts'' in the BAO peak position due to non-linear evolution
\citep[see also][for discussion]{CroSco08,PadWhi09}.
Writing the convolution term in
\begin{equation}
  \delta = \delta_L + \int\frac{d^3p}{(2\pi)^3}
  F_2(\mathbf{p},\mathbf{k}-\mathbf{p})
  \delta_L(\mathbf{p})\delta_L(\mathbf{k}-\mathbf{p}) + \cdots
\end{equation}
in configuration space we have for standard perturbation theory
\citep{BJCP92,SheZal12}
\begin{equation}
  \delta = \delta_L + \frac{17}{21}\delta_L^2
         + \mathbf{\Psi}\cdot\mathbf{\nabla}\delta_L
         + \frac{2}{7} T^2 + \cdots
\end{equation}
where $T$ represent (traceless) shear terms and the
$\mathbf{\Psi}\cdot\mathbf{\nabla}\delta$
term (from the $\mathbf{p}_1\cdot\mathbf{p}_2$ term in
Eq.~\ref{eqn:F2spt}) is largely responsible for the shift of the peak.
In the Zeldovich approximation the expansion to second order is
\begin{equation}
  \delta = \delta_L + \frac{2}{3}\delta_L^2
         + \mathbf{\Psi}\cdot\mathbf{\nabla}\delta_L
         + \frac{1}{2} T^2 + \cdots \quad .
\end{equation}
Note that the shift term is the same, but the growth and shear/anisotropy
terms are slightly different (these terms match if we include the $2^{\rm nd}$
order Lagrangian kernel, i.e.~use 2LPT rather than Zeldovich).
This suggests that the Zeldovich approximation should approximately predict
the small shift in the acoustic peak due to mode-coupling as structure goes
non-linear and the dimunition of this effect due to reconstruction.
Further discussion and comparison of Eulerian and Lagrangian theories in the
special case of one spatial dimension can be found in \citet{McQWhi15}.

\end{appendix}

\label{lastpage}
\end{document}